\documentclass[a4paper]{spie}  

\usepackage{amsmath,amsfonts,amssymb}
\usepackage{graphicx}
\usepackage[colorlinks=true, allcolors=blue]{hyperref}

\usepackage{siunitx}
\DeclareSIUnit{\ph}{photon}
\DeclareSIUnit{\pix}{pixel}
\DeclareSIUnit{\tarcsecond}{arcsecond}
\DeclareSIUnit{\el}{\elementarycharge\tothe{-}}
\DeclareSIUnit{\psb}{\ph\per\second\per\Square\metre\per\Square\tarcsecond\per\micro\metre}
\DeclareSIUnit{\micron}{\micro\metre}

\title{The Australian Space Eye: studying the history of galaxy formation with a CubeSat}

\author[a]{Anthony Horton}
\author[a,b]{Lee Spitler}
\author[c]{Naomi Mathers}
\author[c]{Michael Petkovic}
\author[d]{Douglas Griffin}
\author[d]{Simon Barraclough}
\author[d]{Craig Benson}
\author[d]{Igor Dimitrijevic}
\author[d]{Andrew Lambert}
\author[e]{Anthony Previte}
\author[e]{John Bowen}
\author[e]{Solomon Westerman}
\author[e,f]{Jordi Puig-Suari}
\author[b]{Sam Reisenfeld}
\author[a]{Jon Lawrence}
\author[a]{Ross Zhelem}
\author[c]{Matthew Colless}
\author[d]{Russell Boyce}
\affil[a]{Australian Astronomical Observatory, Sydney, Australia}
\affil[b]{Macquarie University, Sydney, Australia}
\affil[c]{Australian National University, Canberra, Australia}
\affil[d]{UNSW Canberra, Canberra, Australia}
\affil[e]{Tyvak Inc., Irvine CA, USA}
\affil[f]{Cal Poly, San Luis Obispo CA, USA}

\authorinfo{Send correspondence to A.J.H: E-mail: anthony.horton@aao.gov.au, Telephone: +61 2 9372 4847}

\begin{document} 
\maketitle

\begin{abstract}

The Australian Space Eye is a proposed astronomical telescope based on a \SI{6}{U} CubeSat platform. The Space Eye will exploit the low level of systematic errors achievable with a small space based telescope to enable high accuracy measurements of the optical extragalactic background light and low surface brightness emission around nearby galaxies. This project is also a demonstrator for several technologies with general applicability to astronomical observations from nanosatellites. Space Eye is based around a 90 mm aperture clear aperture all refractive telescope for broadband wide field imaging in the $i'$ and $z'$ bands. 

\end{abstract}


\section{INTRODUCTION}
\label{sec:intro}  

In December 2014 the Advanced Instrumentation Technology Centre at the Australian National University hosted the AstroSats 2014 workshop. The workshop was held to explore concepts for astronomical nanosatellite missions, with the goal of identifying viable missions that could be funded within existing Australian grant schemes. Concepts were required to be justifiable by the expected scientific results alone, i.e.\ without reference to the development of technology or accumulation of expertise, valuable as those outcomes would be. Additionally proposed spacecraft were to be no larger than a \SI{6}{U} CubeSat. The Australian Space Eye was conceived in response to this call for proposals.

Meeting the combination of scientific value, cost and size constraints is very challenging, essentially it means identifying a compelling science programme that is within the capabilities of a very small instrument in low earth orbit but which could not be done with (much larger) ground based instruments for comparable cost.  The two most well known benefits of situating an astronomical telescope above the Earth's atmosphere are accessing wavelengths that are absorbed by the Earth's atmosphere and avoiding the degradation of spatial resolution caused by atmospheric turbulence. Neither of these are of great relevance to an astronomical CubeSat however, at least one conceived within the AstroSats constraints. The gamma ray, X-ray, far ultraviolet and infrared wavelength regions all require exotic optics and/or image sensors which are not compatible with the cost, volume, thermal and power limitations of the platform. For this reason we limited our consideration to the near ultravoilet to very near infrared wavelength region ($\sim200$--\SI{1000}{\nano\metre}) that is accessible with silicon based image sensors and conventional optics. Within this wavelength range there is no spatial resolution advantage to being above the Earth's atmosphere either, the fundamental diffraction limit of a CubeSat sized telescope aperture ($\lesssim\SI{90}{\milli\metre}$ diameter) is greater than atmospheric seeing at the same wavelengths.
 
There are however two other effects of the atmosphere which are relevant to a CubeSat based optical telescope: atmospheric scattering and emission.  Atmospheric scattering spreads the light from astronomical objects in a similar way to scattering from the instrument's optics however the impact is compounded by the fact that the distribution of aerosols in the atmosphere is spatially and temporally variable, the amount of scattered light surrounding a given source can vary by several percent of the source brightness even in good `photometric' conditions, on timescales of minutes to months. The variability in the scattering makes it difficult to accurately characterise and subtract \cite{McGraw2010}, thereby introducing problematic systematic errors.  Similar issues arise from atmospheric emission, particularly at longer wavelengths ($>\SI{700}{\nano\metre}$). Here the atmosphere glows increasingly brightly, primarily due to line emission from OH$^*$ molecules in the mesosphere, reducing the sensitivity of ground based telescopes. This emission is sensitive to dynamic processes in the upper atmosphere (e.g.\ gravity waves) and consequenty it is also spatially and temporally variable on timescales of minutes and longer \cite{Moreels2008} which makes accurate subtraction difficult.  As a result of these effects locating a telescope in space not only helpfully reduces both scattered light and sky background levels but crucially makes both far more stable.  Above the atmosphere the only sources of scattered light are associated with the instrument itself and the dominant source of sky background is the zodiacal light \cite{Leinert1998} which, while it does exhibit large scale spatial and seasonal variations, is much less variable, more uniform and more predictable than atmospheric emission.

The scientific competitiveness of small telescope for certain types of observations is well proven. For example while large astronomical telescopes are able to detect extremely faint compact or point-like objects their sensitivity to diffuse emission is limited by systematic errors from a number of sources.  A significant contribution to these systematic errors is contamination with light from brighter objects within/near the instrument's field of view, caused by a combination of diffraction, scattered light and internal reflections. The difficulties caused by these effects have been discussed by, for example, Sandin \cite{Sandin2014,Sandin2015} and by Duc et al \cite{Duc2014}, who noted that the simpler optics of small telescopes suffer less from internal reflections. In addition small telescopes can more readily be constructed with unobscured apertures to minimise diffraction, and simpler optics also make them less prone to internal scattering of light. The resulting competitiveness of optimised small telescopes in the context of `low surface brightness' (LSB) imaging has been demonstrated by the Dragonfly Telephoto Array, an astronomical imaging system based on an array of telephoto camera lenses \cite{Abraham2014}.  This instrument exhibits less diffraction/scattering/internal reflections than other telescopes for which comparable data are available \cite{Abraham2014,Sandin2014} and the system has produced impressive results, including the discovery of a new class of `ultra-diffuse galaxies'\cite{VanDokkum2015}.  Two of this paper's authors (AJH \&\ LS) are currently assembling a ground based instrument based on the same principles, the Huntsman Telephoto Array\cite{Horton2016}.

We conclude that a CubeSat based astronomical telescope may be scientifically competitive, especially an optimised telescope performing measurements that are typically limited by systematic errors. We identify the 700--\SI{1000}{\nano\metre} wavelength region as particularly promising as it is accessible to instrumentation compatible with the constraints of a CubeSat mission while ground based instruments would be hampered by bright atmospheric emission, in terms of both reduced sensitivity and increased systematic errors.  We have based the Australian Space Eye proposal on two science goals which would be well served by a telescope operating in this regime, the measurement of extragalactic background light and imaging of low surface brightness structures around nearby galaxies.  These are discussed in more detail in the next section.

\section{SCIENCE CASE}
\label{sec:aims}
\subsection{Extragalactic Background Light}

A fundamental measurement of the universe is its total luminosity, which contains the entire history of all sources of light, from signatures of the Big Bang in the cosmic microwave background and, at optical wavelengths, all past radiative processes including light from stars and even black hole accretion disks. Although the microwave background has been securely measured the cosmic optical background (the total luminosity at optical wavelengths) has not due to our non-ideal vantage point - we are embedded in the dust cloud of the solar system. This dust scatters sunlight into the observer's telescope and so we must separate the background light we want to measure from the light scattered by this foreground dust, the so-called Zodiacal light. This has proved very challenging (see Cooray 2016\cite{Cooray2016}), though there have been a few attempts (e.g.\ Bernstein 2007\cite{Bernstein2007}, Aharonian et al.\ 2006\cite{Aharonian2006} and Tsumura et al.\ 2010\cite{Tsumura2010}). What is needed is an instrument that can measure and separate the Zodiacal and extragalatic background light.

One approach for disentangling the Zodiacal and extragalactic light exploits spectral features of the former. As Zodiacal light is scattered sunlight its spectrum is similar to that of the Sun, subject to reddening due to the wavelength dependence of the scattering strength\cite{Giavalisco2002,Aldering2001,Leinert1998}. This includes the strong Calcium triplet absorption lines near \SI{860}{\nano\metre}, which will be present in the Zodiacal light spectrum with the same equivalent width as in the solar spectrum but will be absent from the spectrum of the extragalactic background due to the effects of cosmological redshifts. Consequently the observed equivalent width of absorption lines in the total background light (Zodiacal plus extragalactic) provided a measure of the relative contributions of the two components. This was the goal of one of the instruments of the CIBER sounding rocket experiment\cite{Zemcov2013}, a narrowband imaging spectrometer based on a tilted objective filter\cite{Korngut2013}, however the measurement proved extremely difficult to make with the limited total exposure time of a sounding rocket campaign. An orbital space telescope, able to accumulate a year or more of observing time, promises secure measurements of both the absolute total sky brightness and the relative contributions of the Zodiacal light and extragalactic background.

\subsection{Low Surface Brightness Galaxies}

A new ground-based imaging system called the Dragonfly Telephoto Array\cite{Abraham2014} uses commercial-off-the-shelf camera lenses to reach $16\times$ (3 magnitudes) fainter surface brightness levels than existing telescopes at optical wavelengths (0.4--\SI{0.7}{\micron}, $g'$ and $r'$ bands). The key innovation is the use of unobscured refracting optics, which significantly reduce scattered and diffracted light compared to conventional large reflecting telescopes and bring the limiting systematic uncertainties down to \SI{8e-21}{\watt\per\square\meter\per\micron\per\square\tarcsecond} (32 AB mag./arcsecond$^2$ in $g'$)\cite{Sandin2014}.  The Dragonfly observing system has led to a number of important results, including finding an entirely new class of galaxy (e.g.\ van Dokkum, Abraham, Merritt 2014\cite{VanDokkum2014}, van Dokkum et al.\ 2015\cite{VanDokkum2015}). 

By adapting the Dragonfly concept to observing galaxies at longer wavelengths of light from space we can constrain the stellar population properties of these galaxies. The Australian Space Eye will target several nearby galaxies in order to obtain images at the lowest possible surface brightness levels and detect extremely faint structures in their outskirts. When the Space Eye $i'$ and $z'$ band imaging is combined with optical $g'$ and $r'$ band imaging from the Dragonfly and Huntsman\cite{Horton2016} ground-based observing facilities we will have valuable information about the stellar age and chemical content of these faint structures.

\section{SPACE EYE CONCEPT}

\begin{figure}[tp]
  \includegraphics[width=\textwidth]{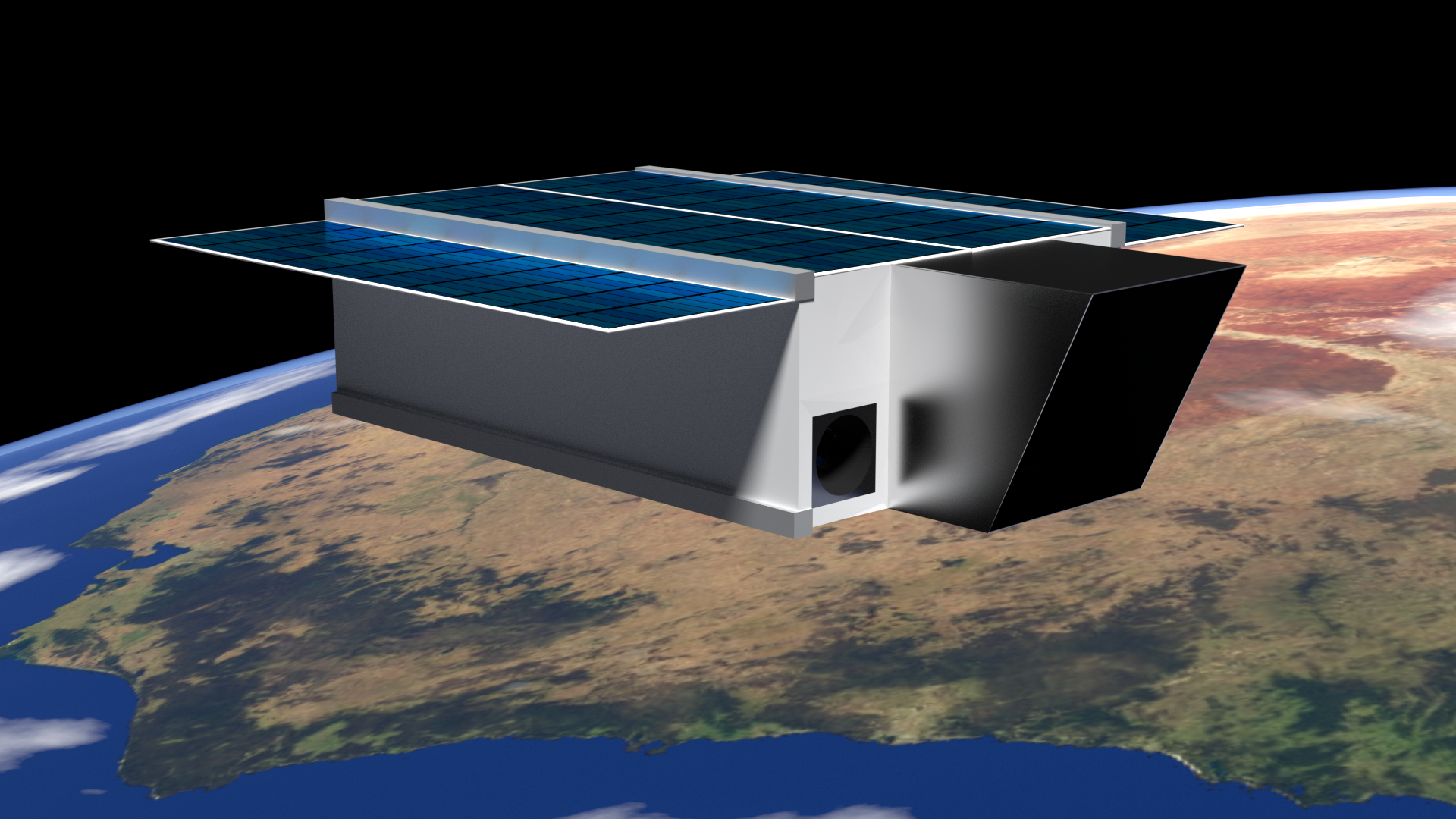}
  \caption{\label{fig:render}Artist's impression of the Australian Space Eye in orbit over the Tasman Sea.}
\end{figure}

The Australian Space Eye is a nano-satellite based on the \SI{6}{U} CubeSat form factor standard\cite{Hevner2011} housing an optical imaging telescope. The specifications of the system are driven by the aims discussed in Sec.~\ref{sec:aims}, subject to the constraints discussed in Sec.~\ref{sec:intro}. An artist's impression of Space Eye is shown in Fig.~\ref{fig:render}.

\subsection{Optical Payload}

\subsubsection{Requirements}
\label{sec:payloadreqs}

The scientific goals described in Sec.~\ref{sec:aims} call for a wide field imaging instrument with moderate spatial resolution, an exceptionally `clean' and stable point spread function (PSF), and the highest low light sensitivity achievable within the other constraints of the platform. The imager must be capable of both broadband imaging in the $i'$ and $z'$ bands (approximately 700--\SI{850}{\nano\metre} and 850--\SI{1000}{\nano\metre} respectively) as well as measurement of the strength of the Calcium absorption lines in the sky background spectrum.

The optical system is expected to occupy approximately 50\% of the spacecraft internal volume. The optical axis will be aligned with the long axis of the spacecraft body and positioned on the centreline to assist with centre of mass positioning. The bulk of the optical payload is therefore confined to an approximately $300 \times 100 \times \SI{100}{\milli\metre}$ \SI{3}{U} volume, however 1--\SI{2}{U} of the remaining \SI{3}{U} of internal volume will be available for payload electronics. The largest practical optical aperture within these constraints is $\sim\SI{90}{\milli\metre}$ in diameter.

Considerations of cosmic variance, sky background gradients and the angular extents of galaxy groups/clusters result in a desire for a field of view at least \SI{1}{\degree} across, preferably close to \SI{2}{\degree}.

We have selected a image scale of \SI{3}{\arcsecond\per\pix}, this is close to Nyquist sampling of the diffraction limited PSF at the Space Eye aperture size and operating wavelengths ($1.22 \lambda / D = \SI{2.0}{\arcsecond}$--\SI{2.8}{\arcsecond}). Coarser spatial sampling could increase the signal to noise ratio for diffuse sources however we are wary of significantly undersampling the PSF due to the potential impact on the accuracy of PSF fitting and subsequent point source subtraction. We will show that Space Eye can still be sky background noise limited at this pixel scale. \SI{3}{\arcsecond\per\pix} is also a good match to the pixel scale of ground based low surface brightness imaging facilities such as the Dragonfly Telephoto Array\cite{Abraham2014} and Huntsman Telephoto Array\cite{Horton2016}.

\begin{table}[bp]
\caption{Summary of main optical payload requirements} 
\label{tab:payload}
\begin{center}       
\begin{tabular}{|l|c|} 
\hline
\rule[-1ex]{0pt}{3.5ex} Field of view & $>\SI{1}{\degree}$ (goal $\sim\SI{2}{\degree}$) \\
\hline
\rule[-1ex]{0pt}{3.5ex} Pixel scale & \SI{3}{\arcsecond\per\pix}  \\
\hline
\rule[-1ex]{0pt}{3.5ex} Wavelength range & 700--\SI{1000}{\nano\metre} \\
\hline
\rule[-1ex]{0pt}{3.5ex}  Aperture diameter & \SI{90}{\milli\metre} \\
\hline
\rule[-1ex]{0pt}{3.5ex}  Optical system dimensions & $300 \times 100 \times \SI{100}{\milli\metre}$ max \\
\hline
\rule[-1ex]{0pt}{3.5ex}  Payload electronics volume & $<\SI{2}{U}$ \\
\hline
\end{tabular}
\end{center}
\end{table}

To accomplish the required measurements of both broadband $i'$ and $z'$ surface brightness and Calcium absorption we proposed to use a set of 6 slightly modified $i'$ and $z'$ band filters with band edges positioned around the strong absoprtion line at \SI{854.2}{\nano\metre}. More details are discussed in Sec.~\ref{sec:mosaic}.

For Space Eye the requirement for a clean and stable PSF translates to minimising all sources of stray light, including surface and bulk scattering, internal reflections, stray sunlight/Earthshine/moonlight and diffraction. 

The highest possible sensitivity will be achieved when the optical aperture is as large as it can be within the space constraints of the CubeSat, the optical throughput and quantum efficiency (QE) of the image sensor are close to 100\%, and the total instrumental noise is below the Poisson noise from the sky background light. 

\subsubsection{Image sensor}

\begin{figure}[tp]
  \center
  \includegraphics[width=0.5\textwidth]{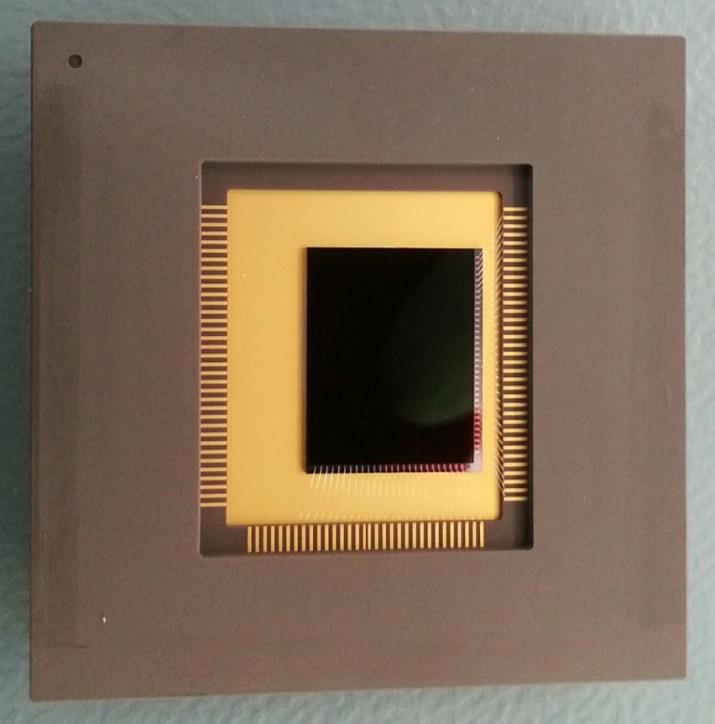}
  \caption{\label{fig:cis115}CIS-115 CMOS image sensor, photo credit e2v.}
\end{figure}

The choice of image sensor is key, not only is it the biggest variable in determining the overall sensitivity of the system but the specifications for many of the other subsystems (optics, data handling, downlink capacity, thermal control, power, etc.) depend on the specifications of the sensor. 

Astronomical instruments operating at these wavelengths typically use charge-coupled device (CCD) image sensors however for Space Eye we have selected the CIS115 CMOS image sensor from e2v\cite{Jorden2014}. The CIS115 is a back-side illuminated image sensor with $2000 \times 1504$ pixels on a \SI{7}{\micron} pitch. The main specifications, taken from the draft e2v datasheet, are given in Tab.~\ref{tab:cis115}.

\begin{table}[bp]
\caption{Summary of CIS115 specifications taken from draft datasheet dated December 2015. All performance values are `typical'.} 
\label{tab:cis115}
\begin{center}       
\begin{tabular}{|l|c|} 
\hline
\rule[-1ex]{0pt}{3.5ex} Number of pixels & $2000 \times 1504$ \\
\hline
\rule[-1ex]{0pt}{3.5ex} Pixel size & \SI{7.0}{\micron} square \\
\hline
\rule[-1ex]{0pt}{3.5ex} Quantum efficiency at \SI{650}{\nano\metre} & \SI{90}{\percent} \\
\hline
\rule[-1ex]{0pt}{3.5ex} Dark current & \SI{20}{\el\per\pix\per\second} at \SI{21}{\celsius} \\
\hline
\rule[-1ex]{0pt}{3.5ex} Read noise & \SI{5}{\el} at \SI{6.2}{\mega\pix\per\second} per channel \\
\hline
\rule[-1ex]{0pt}{3.5ex} Well depth & \SI{27}{\kilo\el\per\pix} (linear), \SI{33}{\kilo\el\per\pix} (saturation) \\
\hline
\rule[-1ex]{0pt}{3.5ex} Non-linearity & $\pm\SI{4}{\percent}$ \\
\hline
\rule[-1ex]{0pt}{3.5ex} Operating temperature & \SI{-55}{\celsius}--+\SI{60}{\celsius} \\
\hline
\rule[-1ex]{0pt}{3.5ex} Power consumption & $\sim\SI{40}{\milli\watt}$ \\
\hline
\end{tabular}
\end{center}
\end{table}

The CIS115 is a new image sensor developed with space based scientific imaging in mind, in particular the ESA JUICE mission. In general its performance is close to that of back side illuminated CCD image sensors operated in inverted mode, the main difference being a 2--$3\times$ higher read noise when compared to the best CCDs. CMOS image sensors offer a number of advantages for a CubeSat space telescope however, including lower power consumption and greater resistance to radiation induced damage than CCDs. Of particular importance to Space Eye is the relatively small pixel size, \SI{7}{\micron} versus the 12--\SI{15}{\micron} of available scientific CCDs. This factor of 2 reduction in pixel size enables a corresponding reduction in the focal length of the optics required for a given on-sky pixel scale which has a significant impact on the design of the optics as discussed in the next section. An additional advantage is the ability to operate regions of the sensor independently of others, opening up the possibility of using small regions for high frame rate autoguiding/star tracking while simultaneously taking a long exposure with the rest of the sensor (`on-chip guiding').

\begin{figure}[tp]
  \center
  \includegraphics[width=0.8\textwidth]{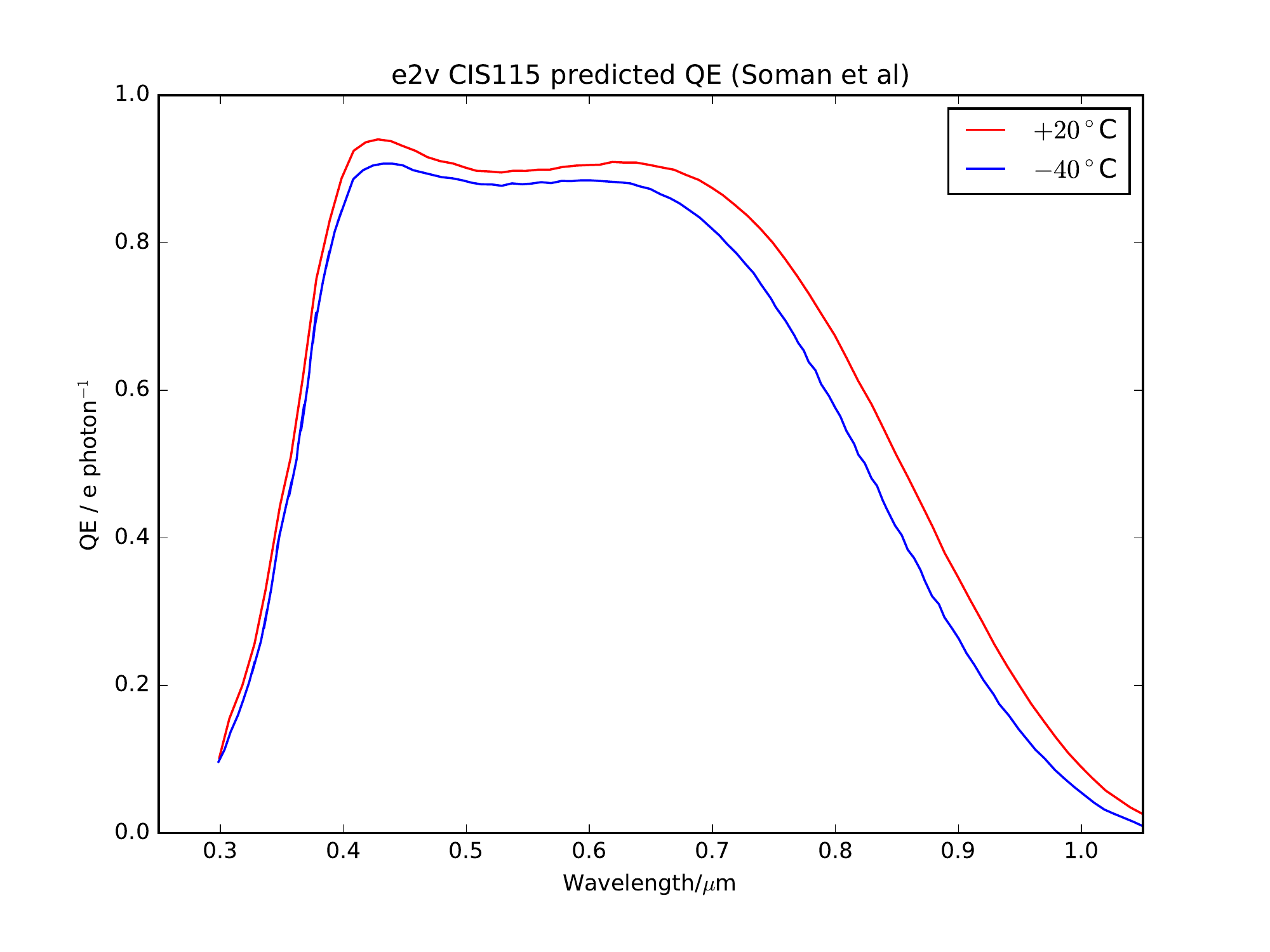}
  \caption{\label{fig:qe}Predicted CIS115 quantum efficiency as a function of wavelength at both \SI{+20}{\celsius} and \SI{-40}{\celsius}, data from Soman et al.\cite{Soman2014}}
\end{figure}

Predicted quantum efficiency data for the CIS115 are plotted in Fig.~\ref{fig:qe}. These data are reproduced from Soman et al.\cite{Soman2014}\ and we regard them as conservative estimates, Wang et al.\cite{Wang2014}\ report slightly higher peak QE from both their own and e2v's measurements of a prototype device (CIS107). The decline in QE for wavelengths greater than \SI{700}{\nano\metre} is due to the increasing transparency of silicon at these wavelengths and is common to all silicon based image sensors. In this wavelength range only deep depletion or high rho CCDs offer significantly higher QE, due to the greater effective thickness of their photosensitive layer. We consider these devices unsuitable for Space Eye however as those currently available have large pixels and, more importantly, require deep cooling ($<\SI{-100}{\celsius}$) to control dark current due to their non-inverted operation. Deep depletion/high rho CMOS image sensors show promise however at the time of writing no devices suitable for Space Eye were known to the authors.

\begin{figure}[tp]
  \center
  \includegraphics[width=0.8\textwidth]{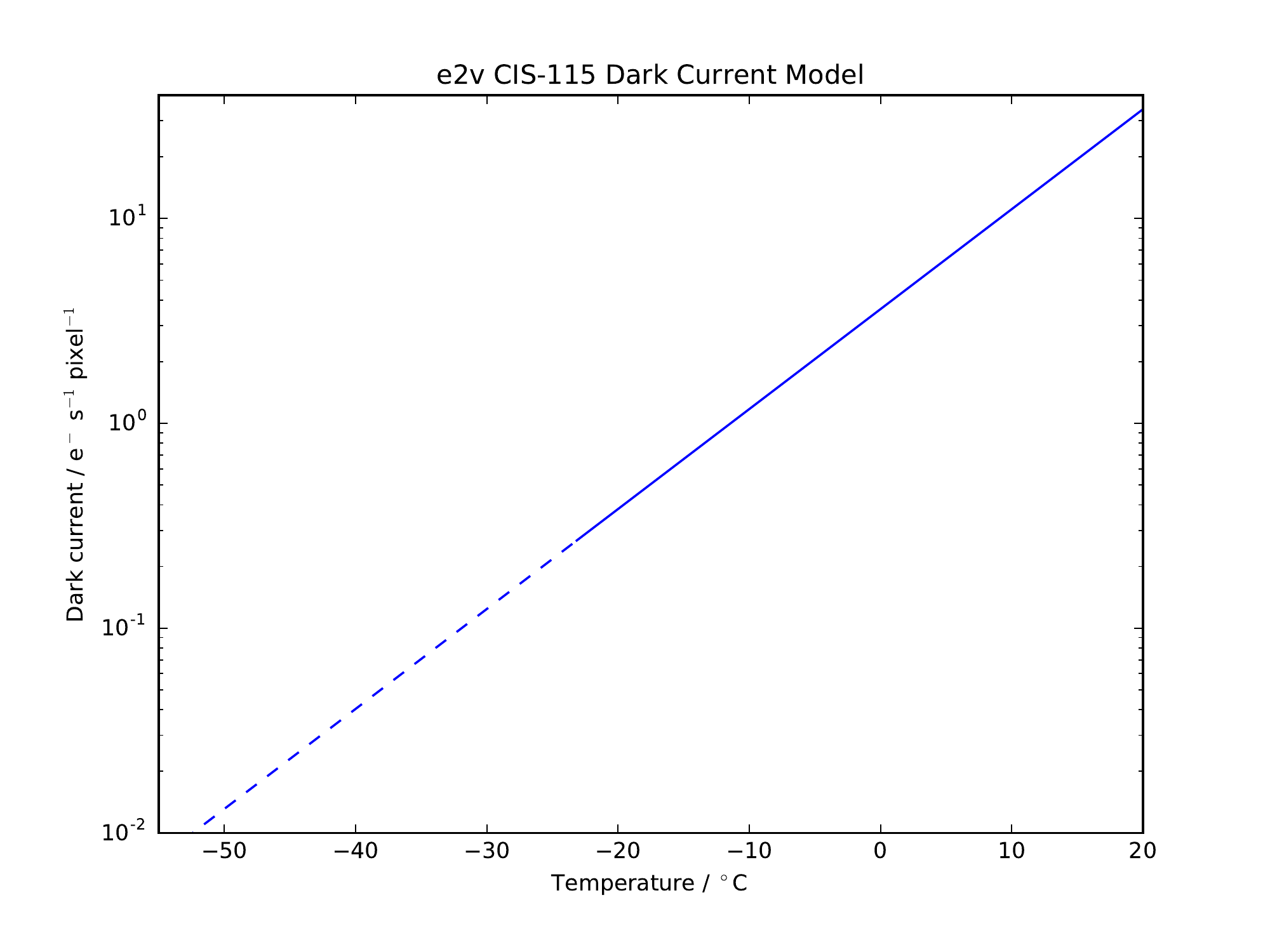}
  \caption{\label{fig:dc}CIS115 dark current as a function of temperature, based on the model from Wang et al.\cite{Wang2014}}
\end{figure}

The dark current of the CIS107 prototype was measured by Wang et al.\ between +\SI{27}{\celsius} and \SI{-23}{\celsius} and they found the expected exponential temperature dependence with a halving of dark current for every \SI{6.2}{\celsius} drop. Their best fit model is shown in Fig.~\ref{fig:dc}. The dark current of cooled production CIS115 image sensors may be somewhat lower, e2v's draft datasheet suggests a typical dark current that is $\sim\SI{50}{\percent}$ lower at \SI{20}{\celsius} that also falls more rapidly with cooling, halving with each 5.5--\SI{6.0}{\celsius} temperature drop. Based on these data we confidently expect a dark current of $\lesssim\SI{0.04}{\el\per\pix\per\second}$ at \SI{-40}{\celsius}. Wang et al.\ also note a helpful reduction in read noise on cooling, to approximately \SI{4}{\el}.

\subsubsection{Telescope optics}

\begin{table}[p]
\caption{Summary of telescope optics specifications} 
\label{tab:optics}
\begin{center}       
\begin{tabular}{|l|c|} 
\hline
\rule[-1ex]{0pt}{3.5ex} Field of view &  $\SI{1.67}{\degree} \times \SI{1.25}{\degree}$ \\
\hline
\rule[-1ex]{0pt}{3.5ex} Effective focal length & \SI{481}{\milli\metre} \\
\hline
\rule[-1ex]{0pt}{3.5ex} Aperture diameter & \SI{90}{\milli\metre} \\
\hline
\rule[-1ex]{0pt}{3.5ex} Focal ratio & $f/5.34$ \\
\hline
\rule[-1ex]{0pt}{3.5ex} Wavelength range & 700--\SI{1000}{\nano\metre} \\
\hline
\rule[-1ex]{0pt}{3.5ex}  Overall length & \SI{250}{\milli\metre} \\
\hline
\end{tabular}
\end{center}
\end{table}

\begin{figure}[p]
  \center
  \includegraphics[width=0.8\textwidth]{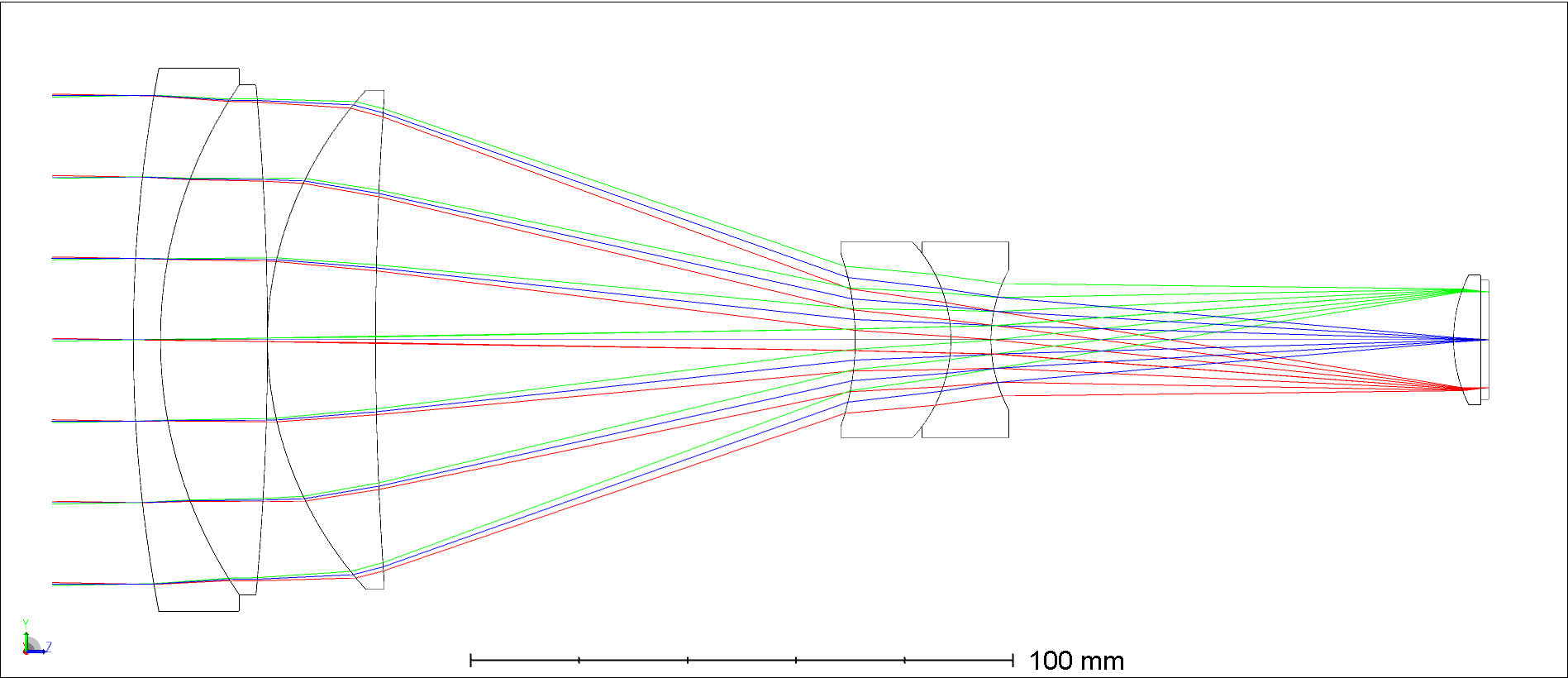}
  \caption{\label{fig:optics}Cross section optical layout diagram of the baseline design for the Australian Space Eye.}
\end{figure}

\begin{figure}[p]
  \center
  \includegraphics[width=0.8\textwidth]{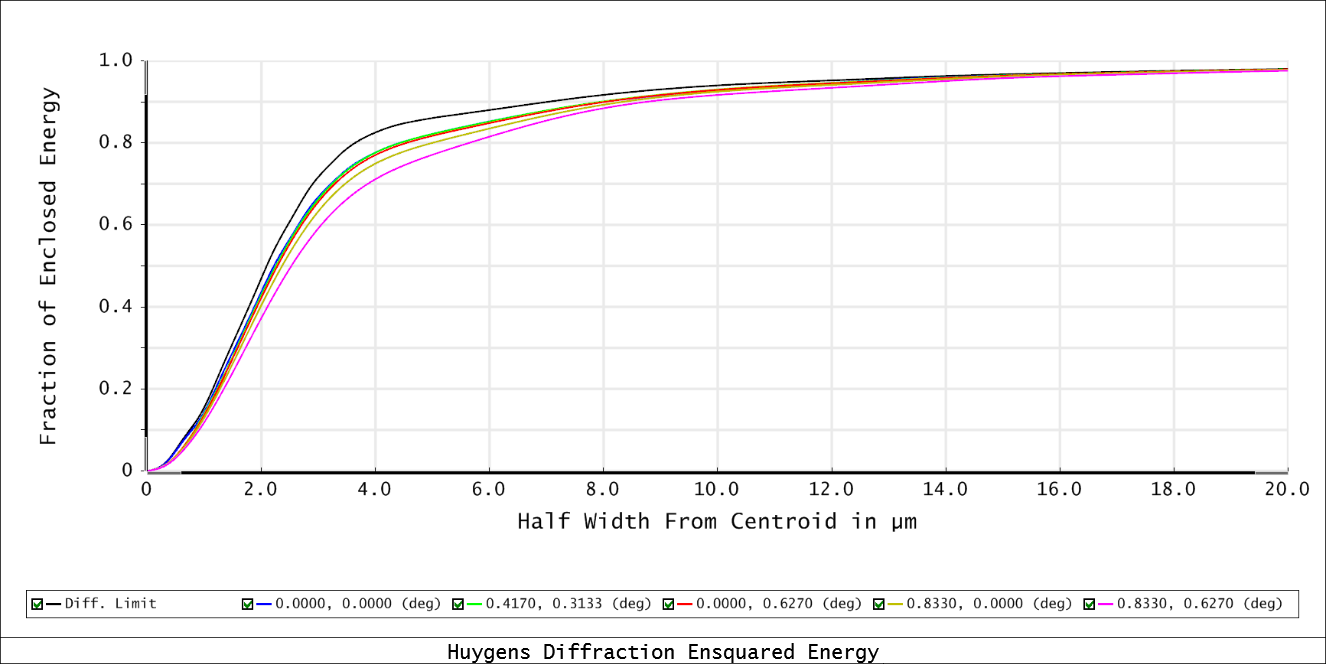}
  \caption{\label{fig:ensquared}Ensquared energy as a function of half-width for the baseline optical design.}
\end{figure}

The combination of the requirements from Tab.~\ref{tab:payload} and the image sensor specifications from Tab.~\ref{tab:cis115} enable us to determine the remaining requirements for the telescope optics. The resulting specifications are summarised in Tab.~\ref{tab:optics}. The biggest consideration for the design of the Space Eye optics is minimising the wings of the PSF due to internal scattering, reflections and diffraction. For this reason we have decided to use an all refractive design. Abraham and van Dokkum\cite{Abraham2014} have argued that refractive designs have a fundamental advantage over reflecting or catadioptric telescopes in this respect, and comparisons of telescope PSF measurements by Sandin\cite{Sandin2014} appear to support this view. Note that the relatively small pixels of the CIS115 image sensor are essential to allow a refractive design to be used. With \SI{3}{\arcsecond\per\pix} and \SI{7}{\micron} pixels the required effective focal length is about twice the length of the space available for the optics, necessitating a moderately telephoto design but not presenting any insuperable problems. With the larger pixels typical of back side illuminated CCDs the focal length would have to be approximately four times the length of the available space, in which case folded light path reflective or catadioptric designs would be the only practical options.

We have produced a baseline optical design for Space Eye which is illustrated in figure \ref{fig:optics}. The design consists of 6 elements in 4 groups with two CaF$_2$ elements (L2 and L3) and two aspheric surfaces (L3-S1 and L4-S1). Six elements were required to achieve the desired image quality within the overall length constraints however we are able to limit the number of vacuum-glass interfaces to 7 which will assist with minimising surface scattering and internal reflections/ghosting. The image quality is essentially diffraction limited across the full field of view, as confirmed by the ensquared energy plots in Fig.~\ref{fig:ensquared}. The calculation was performed using the Huygens-Fresnel method with the Zemax OpticStudio software. The ensquared energies are polychromatic (700--\SI{1000}{\nano\metre}) and have been calculated for 5 field points at a range of positions from the centre to the corner of the field of view.

Bulk scattering will be minimised through the use of high purity, high homogeneity lens blanks. Super polishing and high performance anti-reflection coatings will be used to minimise surface scattering and internal reflections. We are pursuing the possibility of using nano-structured anti-reflection coatings (as employed in some high end DSLR lenses) for this purpose, these are capable of very low reflectivities for a wide range of wavelengths and angles of incidence. Internal knife-edge baffles will be used, along with an external deployable baffle designed to prevent illumination of the optics by sunlight, Earthshine or moonlight during observations. A 700--\SI{1000}{\nano\metre} bandpass filter will be applied to the 1st lens surface. A full stray light analysis will be performed as part of a general review of the optical design when the project is confirmed however we are confident that we will be able to acheive the required clean and stable PSF.

The telescope will include a bistable optical shutter between the L5 and L6 elements to enable on orbit image sensor dark current measurements.

\subsubsection{Mosaic filter}
\label{sec:mosaic}

\begin{figure}[pt]
  \center
  \includegraphics[width=0.8\textwidth]{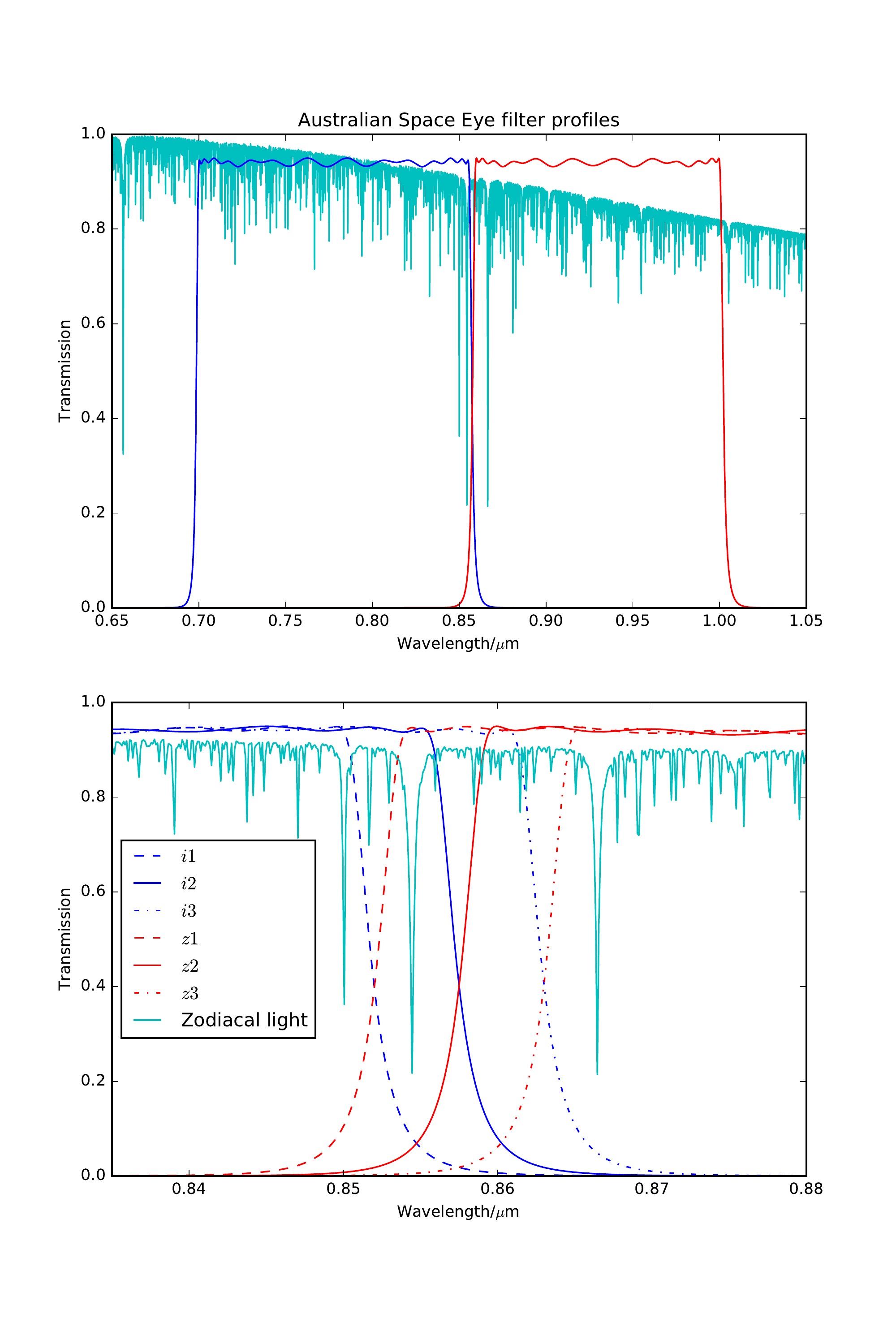}
  \caption{\label{fig:filters}Nominal filter transmission profiles shown together with a model of the Zodiacal light photon spectral flux density.}
\end{figure}

In order to provide the necessary data to allow both broadband $i'$ and $z'$ band imagery and seperate measurements of the Zodiacal light and extragalactic components of the sky background we propose the use of 6 broadband filters, 3 variants on the $i'$ filter and 3 variants on the $z'$ filter. Nominal transmission profiles for these filters are shown in Fig.~\ref{fig:filters}, together with a model of the Zodiacal light photon spectral flux density. The filters differ in their red cutoff wavelength in the case of the $i'$ filters and their blue cutoff wavelength in the case of the $z'$ filters. The cutoffs are chosen to bracket the strongest of the Calcium triplet absorption lines (\SI{854.2}{\nano\metre}) as well as an adjacent region of continuum. The pairwise surface brightness differences between filters provide information on the relative strength of the Calcium absorption that can used together with sky background models to separate the Zodiacal light and extragalactic components. Meanwhile the sum of the 3 $i'$ or $z'$ band filters gives deep broadband data with effective filter response close to the standard $i'$ or $z'$ filter bands.

\begin{figure}[tbp]
  \center
  \includegraphics[width=0.4\textwidth]{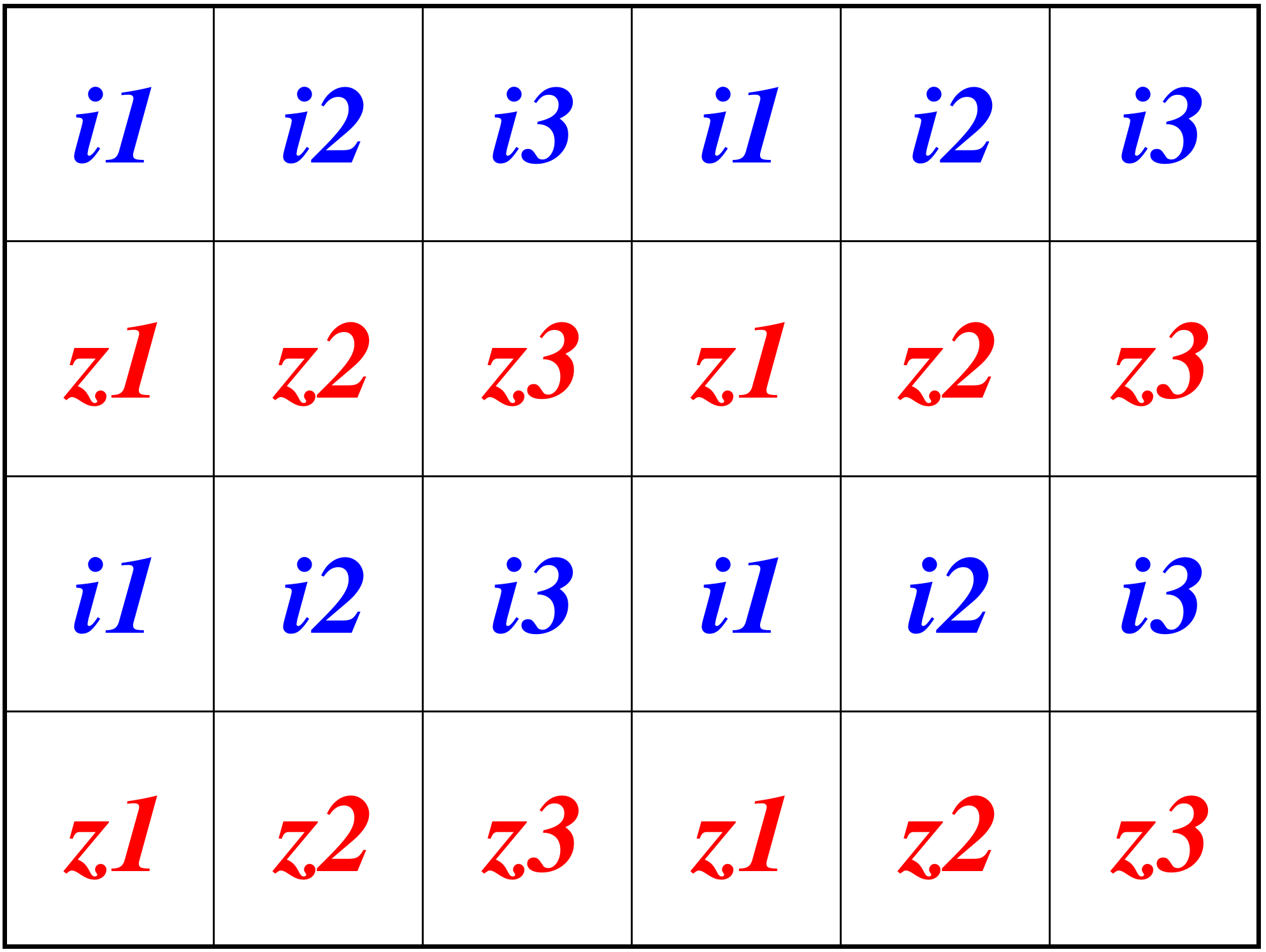}
  \caption{\label{fig:mosaic}Example mosaic filter layout. Each filter covers an area of $333 \times \SI{376}{\pix}$, $\SI{16.7}{\arcminute} \times \SI{18.8}{\arcminute}$ on sky.}
\end{figure}

Multiband astronomical imaging is typically accomplished by taking full frame images through each individual filter sequentially, i.e.\ temporal multiplexing. This approach requires a set of full frame filters and some sort of filter exchange mechanism, e.g.\ a filter wheel. For space based instruments, and especially those intended for nanosatellites, there are strong incentives to avoid introducing mechanisms if at all possible due to complexity, cost and potential for failure. Consequently we favour a spatial multiplexing based approach in which the field of view is divided up amongst the 6 different filters by a fixed mosaic filter at the focal plane, an example layout of which can be seen in Fig.~\ref{fig:mosaic}. By taking sequences of 6 images with pointing offsets equal to the dimensions of the filter regions contiguous images can be built up for each filter. 

\subsubsection{Performance modelling}
\label{sec:perf}

In order to determine appropriate operational parameters and predict the approximate sensitivity of Space Eye we have developed performance models for the optical payload. These are not full end-to-end simulations but are instead parametric models, suitable for efficiently exploring parameter space ahead of the more detailed analysis to follow.

\begin{figure}[tbp]
  \center
  \includegraphics[width=0.8\textwidth]{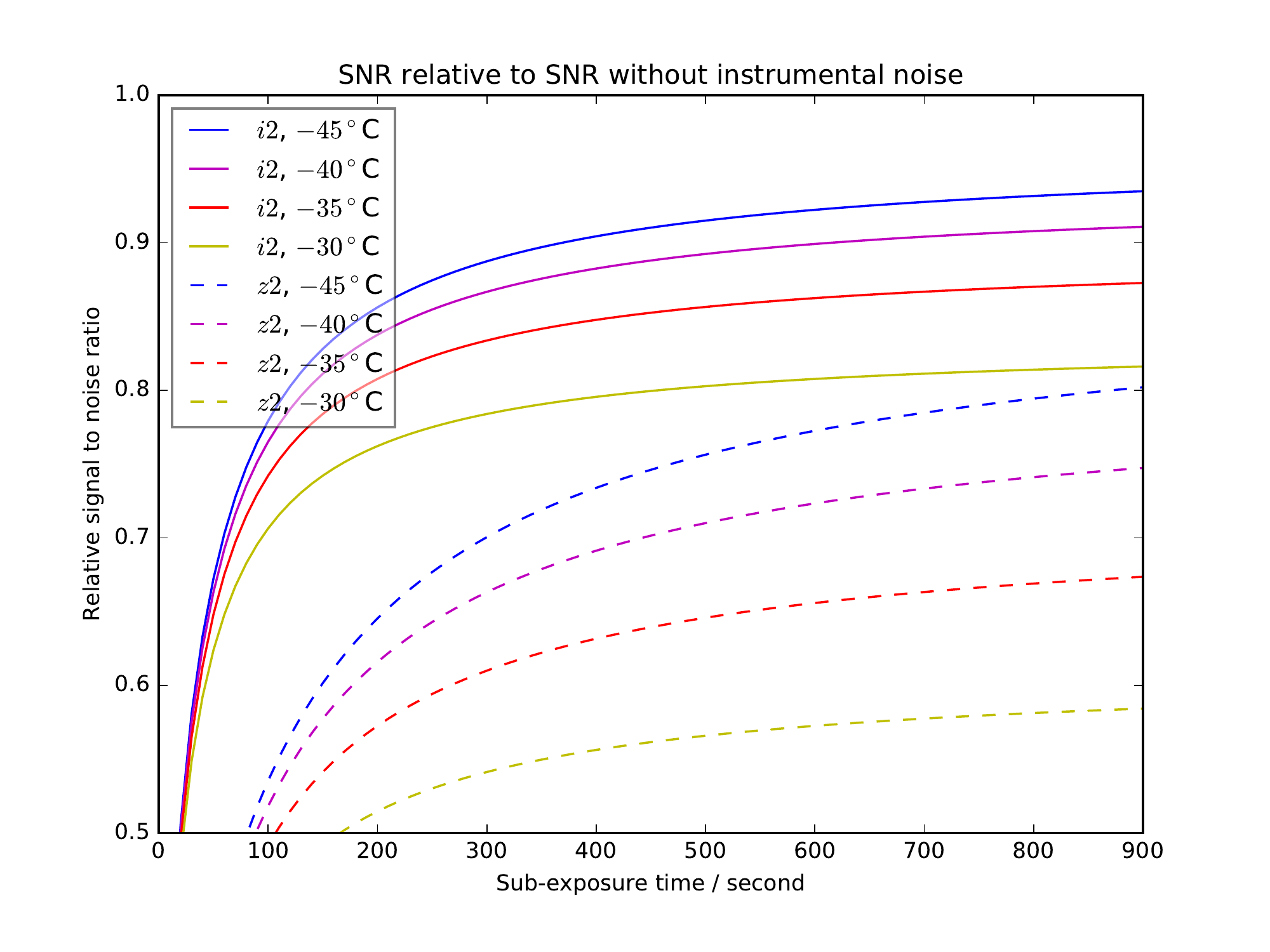}
  \caption{\label{fig:relsnr}Predicted SNR relative to the SNR without instrumental noise for the $i2$ and $z2$ filters as a function sub-exposure time for a range of image sensor temperatures. The calculation uses the ecliptic pole Zodical light surface brightness values.}
\end{figure}

As noted in Sec.~\ref{sec:payloadreqs} obtaining the maximum sensitivty for a fixed telescope aperture size requires both maximising the end-to-end efficiency (optical throughput and image sensor QE) and ensuring the instrumental noise sources do not add significantly to the fundamental Poisson noise of the light received from the sky, which at these wavelengths is predominantly Zodiacal light. For the purpose of these calculations we use a model for the Zodiacal light based on that used by the Hubble Space Telescope Exposure Time Calculator\cite{Giavalisco2002}. The starting point is a solar spectrum from Colina, Bohlin and Castelli\cite{Colina1996}, to which we apply a normalisation, reddening and spatial dependency following the prescription of Leinert et al.\cite{Leinert1998} with the revised parameters from Aldering\cite{Aldering2001}. Using the aperture size, estimated optical throughput, filter transmission profiles, pixel scale and image sensor quantum efficiency we can predict the observed Zodiacal light signal (in photo-electrons per second per pixel) and then, with the estimated image sensor dark current and read noise values, we can predict the signal to noise ratio (SNR). By comparing the predicted SNR with the value it would have in the absence of dark current and read noise we can quantify the degree to which instrumental noise is effecting sensitivity. The key results are summarised in Fig.~\ref{fig:relsnr}, which shows the SNR relative to the SNR without instrumental noise for the $i2$ and $z2$ filters for a range of image sensor temperatures and sub-exposure time. 

The calculation uses the Zodiacal light model surface brightness values for the eclipitc poles, which are close to the minimum possible value and therefore place the most stringent demands on instrumental noise. Due primarily to the falling image sensor QE the $z'$ band is more sensitive to instrumental noise and drives the selection of operating parameters. Based on these results we have selected a nominal exposure time for individual science exposures of \SI{600}{\second} and an image sensor operating temperature of \SI{-40}{\celsius}.

\begin{figure}[tbp]
  \center
  \includegraphics[width=0.8\textwidth]{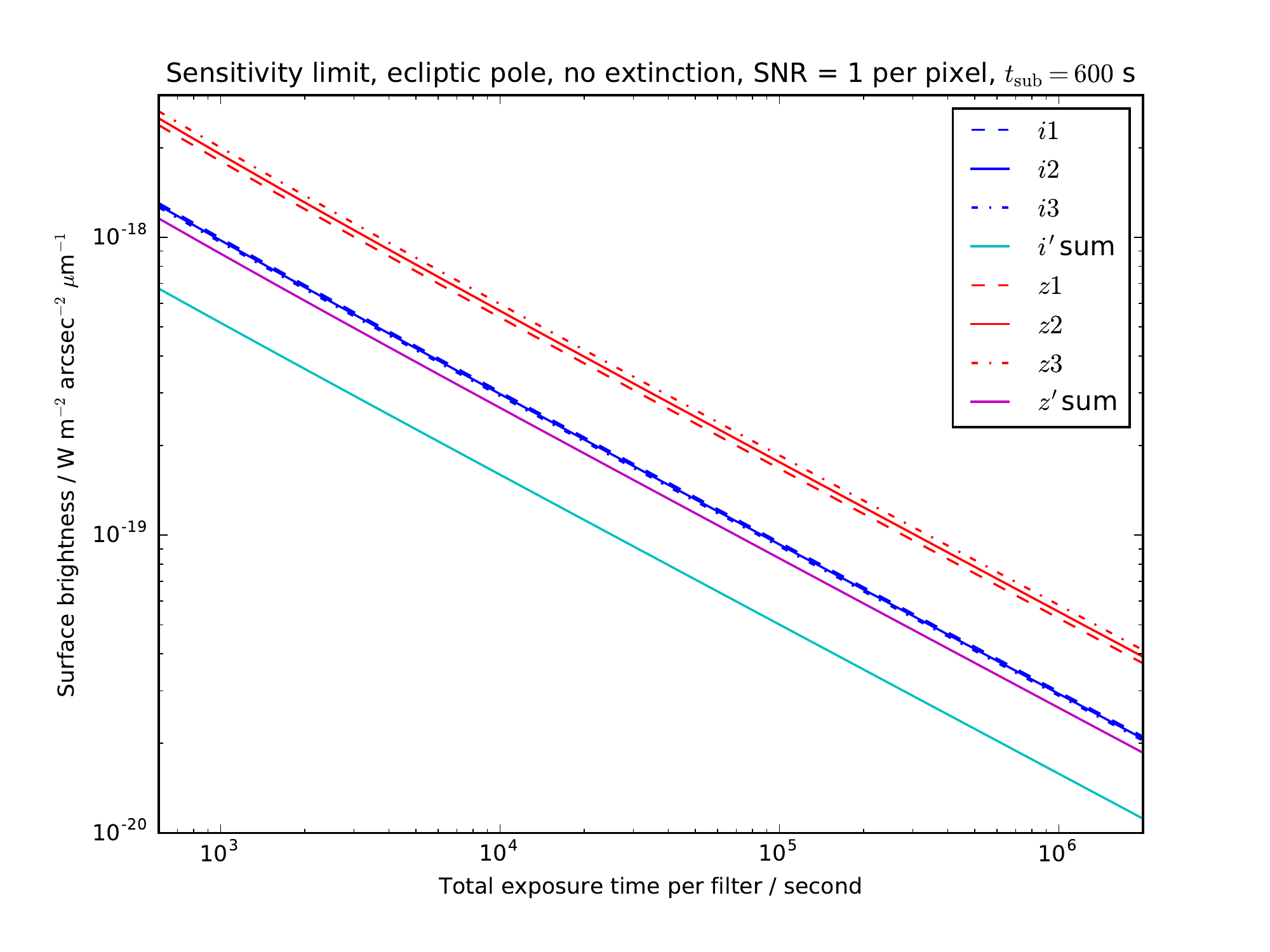}
  \caption{\label{fig:sens}Predicted surface brightness spectral flux density corresponding to a signal to noise ratio of 1 per pixel as a function of total exposure time per filter. The calculcation is for the ecliptic pole Zodiacal light surface brightness, an image sensor temperature of \SI{-40}{\celsius} and sub-exposures of \SI{600}{\second}.}
\end{figure}

Using these parameters we then calculate the predicted sensitivity limit for Space Eye, specifically the source surface brightness spectral flux density that would correspond to a signal to noise of 1 per pixel, for each filter. This is plotted as a function of total exposure time in Fig.~\ref{fig:sens}, with a horizontal scale that runs from \SI{600}{\second} (i.e. a single exposure) to \SI{2e6}{\second}. The latter value corresponds to the approximate total exposure time per filter per celestial hemisphere for a 2 year duration mission, assuming the duty cycle discussed in Sec.~\ref{sec:obsconst}. Of more relevance to the broadband imaging science goals, we have also calculated the predicted sensitivity limit from summing the 3 $i'$ and $z'$ filters. At the end of the 2 year mission we predict ultimate $1\sigma$ per pixel sensitivity limits of \SI{1.1e-20}{\watt\per\metre\squared\per\tarcsecond\squared\per\micron} and \SI{1.9e-20}{\watt\per\metre\squared\per\tarcsecond\squared\per\micron} for $i'$ and $z'$ bands respectively, equivalent to 30.5 and 29.6 in AB magnitude surface brightness units.

\begin{figure}[pt]
  \center
  \includegraphics[width=0.535\textwidth]{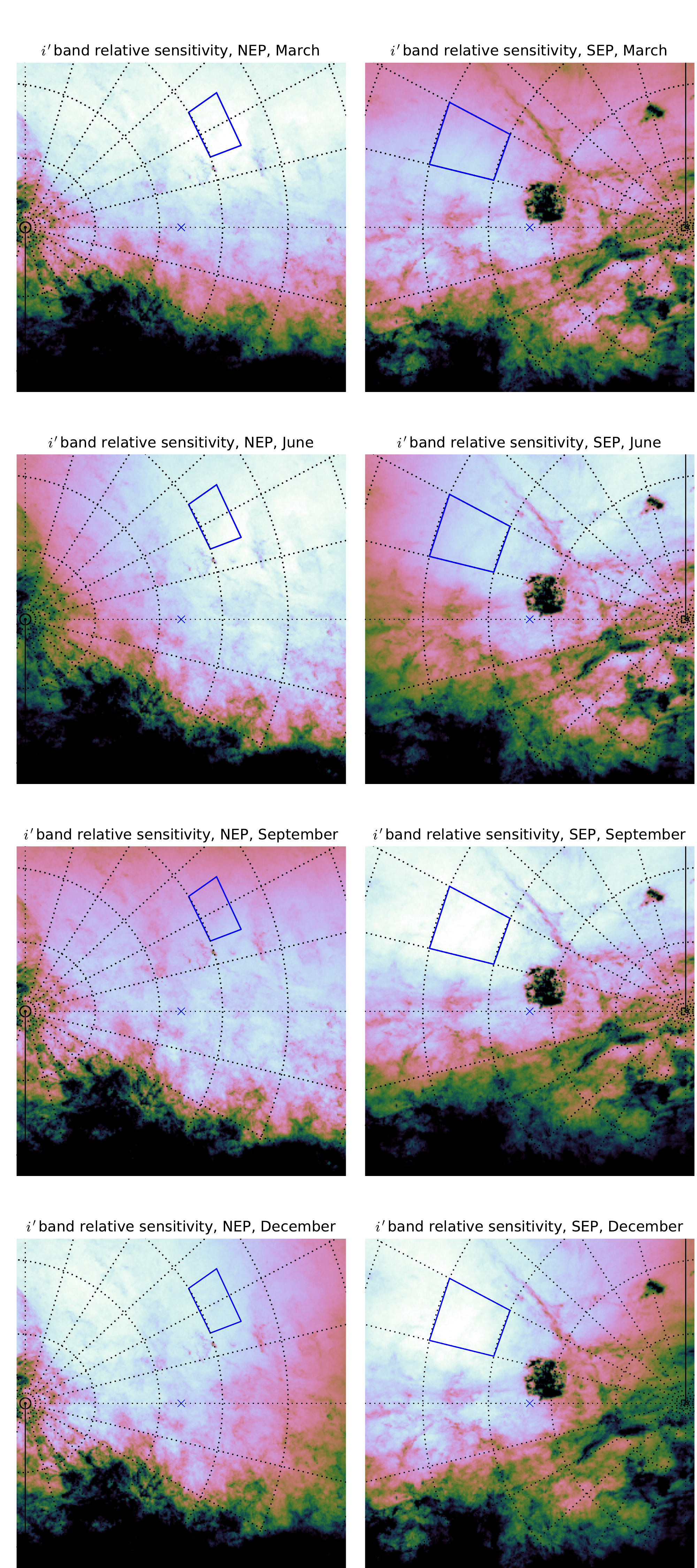}
  \caption{\label{fig:sensmap}Predicted $i'$ band sensitivity relative to the ecliptic pole, zero extinction case. The maps show $\SI{45}{\degree}\times\SI{45}{\degree}$ gnomonic projections centred on the north and south ecliptic poles in March, June, September and December, with a colourmap running from halved sensitivity (black) to full sensitivity (white). Equatorial coordinate grids are also shown, with 1 hour spacing in RA and \SI{10}{\degree} in dec, as well as the outlines of the designated target regions.}
\end{figure}

The Zodiacal light model includes spatial (and seasonal) variations, allowing us to calculate how the predicted sensitivity varies with position on the sky. For extragalactic sources, such as other galaxies and the extragalactic background light, we must also take into account extinction by Galactic dust. For this we use the all sky dust reddening (E(B-V)) maps from the Planck Legacy Archive\cite{Abergel2014} and convert to extinctions at our filter wavlengths using the prescription of Fitzpatrick\cite{Fitzpatrick1999}, with $R=3.1$.  Fig \ref{fig:sensmap} shows the resulting $i'$ band relative sensitivity maps for regions near the ecliptic poles at the times of the equinoxes and solstices. Due to the seasonal variation in Zodiacal light regions close to the north ecliptic pole are preferred from February to July, and close to the south ecliptic pole from August to January. The tightest constraints come the Galactic dust, in order to minimise its effects target should be chosen from within regions between 15.5 and 16.5 hours Right Ascension and +55 and +60 degrees declination in the north, and between 4 and 5 hours Right Ascension and -50 and -60 degrees declination in the south.

\subsection{Spacecraft Bus}

The Space Eye spacecraft bus will be based on the Tyvak Endeavour platform. This highly integrated, high performance platform incorporates almost all the systems required for a functional 3--\SI{12}{U} Cubesat, including Command and Data Handling (C \&\ DH),  Electrical Power System (EPS) and thermal management, Attitude Determination, Control and Navigation System (ADCNS), and structural and mechanical parts.  In a typical configuration the avionics package, including battery modules, occupies approximately \SI{1}{U} of volume.  As noted in Sec.~\ref{sec:payloadreqs} the Space Eye telescope will occupy a \SI{3}{U} volume leaving a final $\sim\SI{2}{U}$ of volume for other mission specific hardware, e.g.\ the image sensor control electronics, image sensor thermal control system (Sec.~\ref{sec:thermal}), image stabilisation system/ADCS 2nd stage (Sec.~\ref{sec:adcs}) and communication equipment.

An initial power budget analysis has indicated that Space Eye will require body mounted solar panels on the largest face of the spacecraft body plus \SI{1}{U} wide deployable panels along both long edges, as shown schematically in figure \ref{fig:render}. Based on a preliminary analysis we believe that standard UHF telemetry/command and S-band downlink systems will be sufficient given the expected data rates (see section \ref{sec:conops})\cite{Reisenfeld2015}.

\subsubsection{Attitude Determination \&\ Control}
\label{sec:adcs}

\begin{figure}[tp]
  \center
  \includegraphics[width=0.7\textwidth]{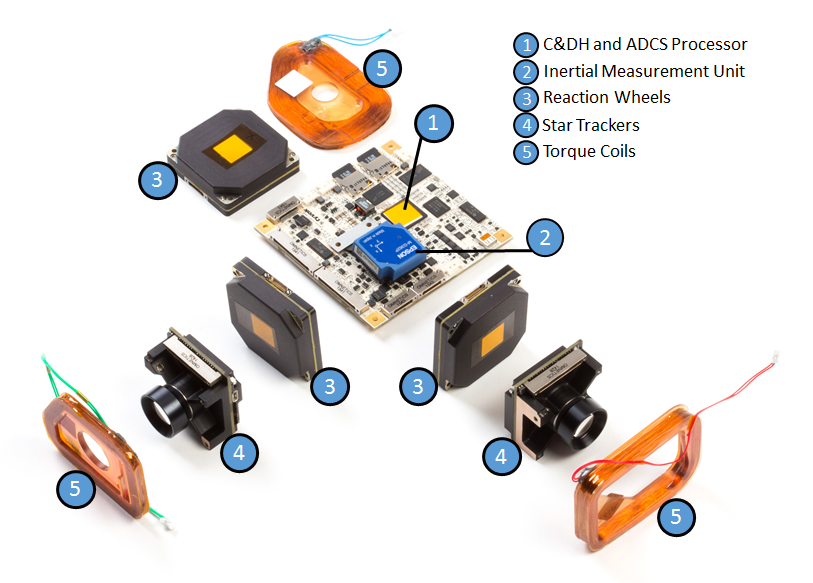}
  \caption{\label{fig:adcs}Main components of the standard Tyvak Endeavour attitude determination and control system (ADCS). Space Eye's ADCS will be based on an updated Endeavour system.}
\end{figure}

The main technical challenge for astronomical imaging from CubeSats is instrument pointing stability. Long exposures are required to prevent image sensor noise overwhelming the faint signals from the sky (see Sec.~\ref{sec:perf}) and the instrument must be kept stable to within less than 1 pixel for the duration of the exposures to avoid blurring.  For Space Eye the required exposure times are \SI{600}{\second} and \SI{1}{\pix} corresponds to \SI{3}{\arcsecond}, these requirements are well beyond the capabilities of any current commercially available CubeSat ADCS system.  Improvements are required in both the attitude determination and attitude control aspects. 

The standard Tyvak Endeavour main attitude determination system is based on a set of two orthogonal Tyvak-developed star trackers and a three-axis MEMS gyro. These sensors are paired with a set of three Tyvak reaction wheels as primary attitude control actuators. The system also incorporates three magnetic torque coils for wheel desaturation and a set of sun sensors and magnetometers to provide coarse attitude determination. Attitude determination and control computations are performed on a dedicated processor on the Endeavor main board. Pointing stability of the Endeavour platform is primarily driven by rate noise from the gyroscope in the attitude control loop. Disturbances from the reaction wheels also contribute jitter to a lesser extent; however, reaction wheel induced jitter can be mitigated with placement and isolation. Modest software and firmware updates to the current Endeavour platform achieve \SI{30}{\arcsecond} (\SI{3}{$\sigma$}) stability in simulation.

Tyvak has investigated several paths to achieve arcsecond level pointing stability with the Endeavour platform. The highest technology readiness level path for arcsecond level pointing of an imaging payload is a two-stage control system with piezoelectric actuators driving lateral translation of the focal plane assembly (FPA) in a high bandwidth second stage. The two-stage approach is well suited to CubeSats since it is orbit agnostic and CubeSat rideshares generally cannot select their orbit. 

The piezoelectric second stage consists of the FPA (image sensor, interface board, mosaic filter and field flatenner lens), piezoelectric actuators driving lateral translation of the FPA, a dedicated processor for calculating guide star centroids and performing control calculations, and the electrical and power system for the piezoelectric actuators. 

Space Eye will use fine star tracking (autoguiding) in the main telescope focal plane to provide the precision pointing information required. Dedicated sensors adjacent to the main image sensor could be used however the CIS115 image sensor and its control electronics are capable of continuously reading out sub-regions of the image sensor for fine star tracking while the remainder of the sensor is simultaneously doing a long science exposure.
A fraction of the science image is lost in this way however avoiding a dedicated set of fine star tracking sensors and associated control electronics reduces complexity considerably.

Preliminary design for the dual-stage controller indicates the control bandwidth of the piezoelectric second stage needs to be greater than \SI{3.5}{\hertz} with centroiding noise of less than \SI{1}{\arcsecond}. Control inputs (centroids from guide stars) need to be computed at roughly \SI{20}{\hertz} and with delay of no more than roughly \SI{1}{\milli\second}. 

Specific piezoelectric actuators have not been analyzed however the required actuator throw is within COTS hardware capability. Significant work remains to refine guide star sensor selection, actuator selection, packaging, and the algorithms to drive the second stage control system.

\subsubsection{Thermal control}
\label{sec:thermal}

\begin{figure}[tbp]
  \begin{center}
  \begin{tabular}{m{0.45\textwidth} m{0.45\textwidth}}
    \includegraphics[width=0.45\textwidth]{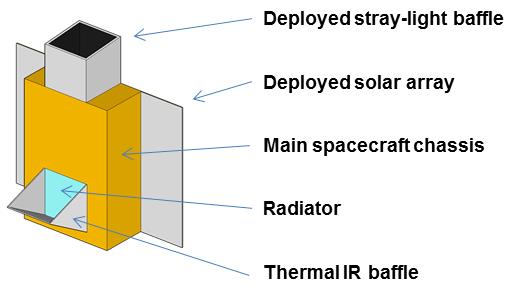} & \includegraphics[width=0.45\textwidth]{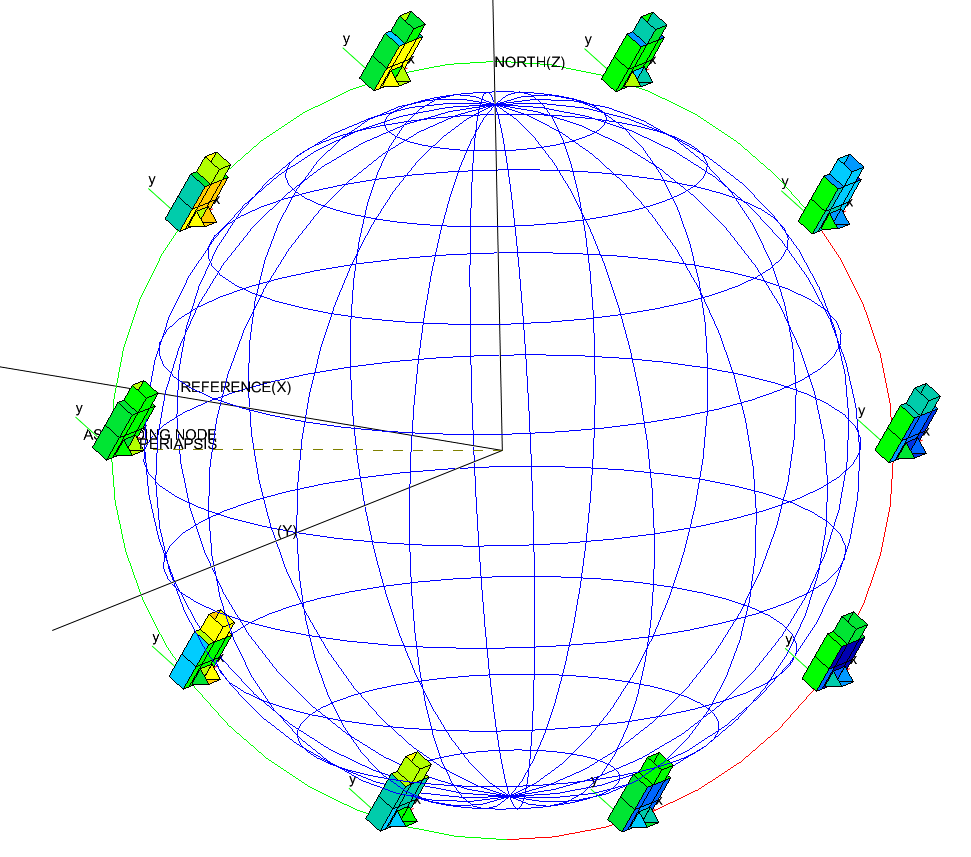}
  \end{tabular}
  \end{center}
  \caption{\label{fig:thermalpics}Configuration of the spacecraft showing the deployed thermal IR radiator baffle (left), and spacecraft attitude and solar illumination during (northern) summer solstice in a 14:00 LTAN orbit without Earth IR avoidance manoeuvres outside of the observation windows (right).}
\end{figure}

\begin{figure}[tbph]
  \center
  \includegraphics[width=0.8\textwidth]{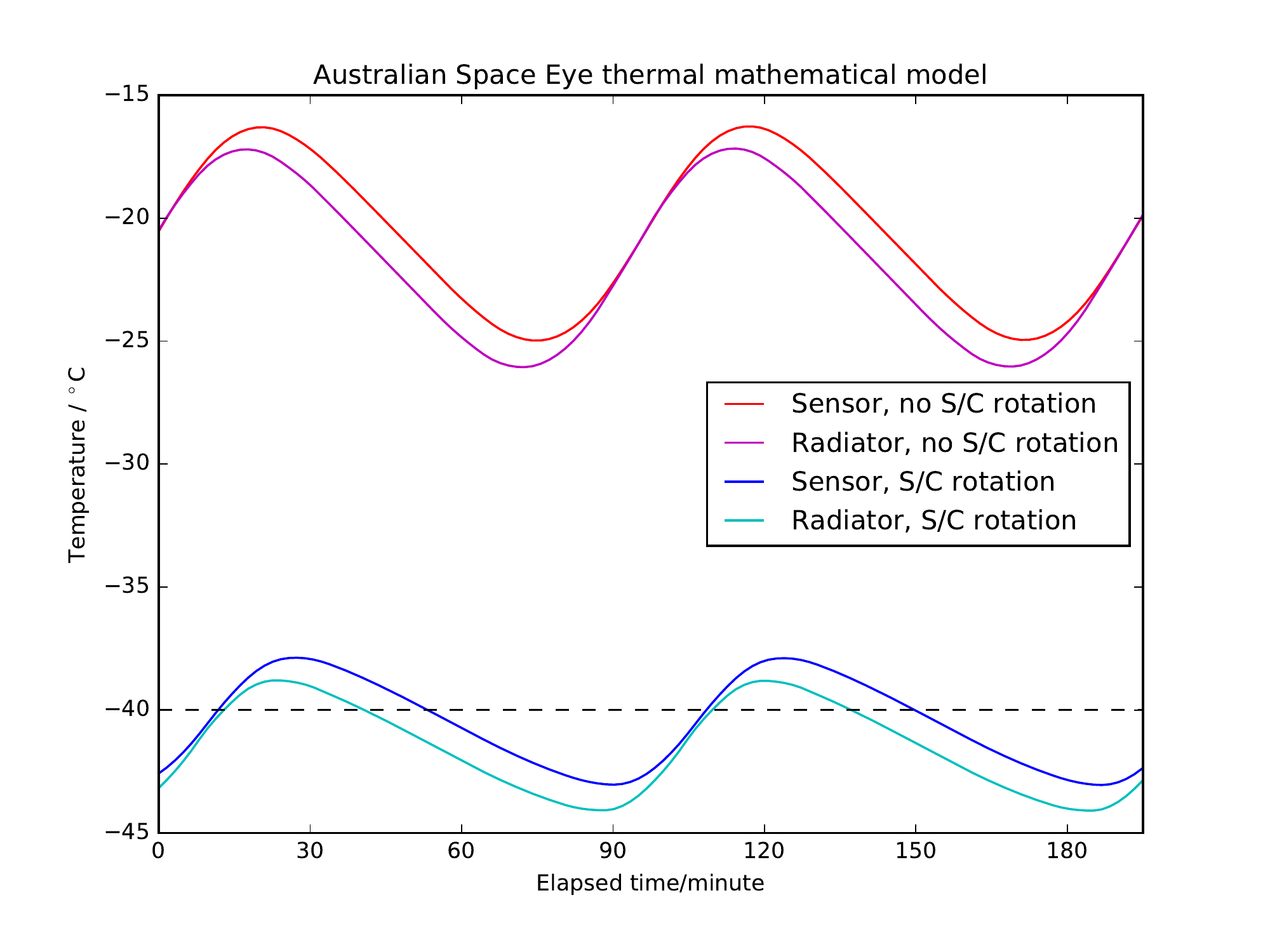}
  \caption{\label{fig:thermalplots} Predicted image sensor and thermal radiator temperature variations during several orbits with a thermal IR baffle, both with and without Earth IR avoidance manoeuvres outside of the observation window. When Earth IR avoidance manoeuvres are used the image sensor temperature remains below the nominal \SI{-40}{\celsius} operating temperature throughout the observation windows.}
\end{figure}

Cooling the image sensor below the temperature where the dark current is a negligible contributor to the system radiometric noise budget (see Sec.~\ref{sec:perf}) is both critical to the scientific performance of the mission as well as a very difficult engineering challenge within the resource constraints of a \SI{6}{U} CubeSat spacecraft. In order to address the feasibility of achieving this requirement, a number of configurations of the spacecraft, attitude control strategies and Concept of Operations need to be assessed. 

In order to cool the detector over an extended period of time a radiator needs to be integrated onto an external face of the spacecraft to reject heat to deep space. Through the orbit the radiator will in general be in radiative exchange with deep space, the Earth, the solar heat load from the direct view of the radiator to the Sun as well as the solar energy reflected from the Earth to the radiator (i.e.\ the Albedo heat load). The overall efficacy of a particular radiator design will then depend on the area, thermal conductivity and thermo-optical properties of the radiator as well as the relative geometry of the radiator surface to the Earth, Sun and any other appendages of the spacecraft. Ultimately, these geometric factors depend on the accommodation of the subsystems within the spacecraft, the attitude of the spacecraft, the season and the orbital parameters of the spacecraft and the location of the spacecraft in its orbit. 

The most favorable orbit, from the point of view of thermal control of the detector radiator, is dawn-dusk Sun synchronous. In this orbit the detector radiator can be placed on the opposite side of the spacecraft to the main solar array. has a very good view factor towards deep space and is protected from direct solar illumination. There are several drawbacks of baselining this orbit which makes it unattractive from the mission level perspective. Firstly, only a relatively small minority of CubeSat launches are dawn-dusk and therefore adopting it involves a risk of incurring a significant programmatic delay to wait for a suitable launch opportunity. Secondly, due to the fact that the sequencing of the deployments of the primary and secondary payloads is planned to optimize the orbital parameters of the primary payload, there is generally an injection error of the CubeSat with respect to a fully sun synchronous orbit. Finally, without a propulsion system, as the orbit decays it will drift further and further away from the ideal orbit and compromise the thermal performance. In the light of these considerations it was decided to abandon the dawn-dusk sun-synchronous orbit approach and consider Sun-synchronous orbits with a Local time of the Ascending Node (LTAN) closer to local noon.

A Thermal Mathematical Model (TMM) of the spacecraft and orbit was built using ESATAN TMS. For the purpose of the analysis we assumed a \SI{600}{\kilo\metre} Sun-synchronous circular orbit with a 96.7 minute period and LTAN of 14:00. The initial analysis showed that a $\SI{100}{\milli\metre} \times \SI{100}{\milli\metre}$ radiator on the anti-Sun facing surface of the spacecraft would not provide sufficient cooling to the image sensor due to the thermal IR load from the Earth (and to a lesser extent the solar albedo load). In order to improve the efficiency of the radiator a deployable Thermal IR baffle was modelled (see Fig.~\ref{fig:thermalpics}). The inside faces of this baffle are to be coated with a low-emissivity coating (for example, vacuum deposited aluminum) to improve the effective view factor of the radiator to deep space while blocking views to Earth. In order to avoid interference with the spacecraft Attitude Determination and Control System (ADCS) the stiffness and first eigenmode of the deployed baffle needs to be outside of the control bandwidth of the fine pointing system.

The TMM indicated that although the thermal IR baffle improves the performance of the radiator, there are periods of the orbit where the radiator is exposed to the Earth and does not provide enough cooling to the image sensor. Fortunately, these periods occur outside the observation windows and therefore the spacecraft is able to slew the radiator away from the Earth without significantly compromising the power generating capacity of the solar arrays. This can be seen in Fig.~\ref{fig:thermalplots}, where the image sensor temperatures are plotted during several orbits with and without the Earth IR avoidance manoeuvres. The design and analysis carried out on the image sensor cooling system shows the fundamental feasibility for this aspect of the mission. 

\section{Concept of Operations}
\label{sec:conops}

The concept of operations for Space Eye is very much a work in progress at the time of writing, however some aspects have been worked out. The science aims of the project require the telescope to obtain repeated long exposure images of a small number of target fields for as long as possible.

\subsection{Target fields}

The exact positions of the target fields will be chosen based on a number of factors, including the positions of suitable guide stars, avoidance of very bright stars and the locations of galaxies and galaxy groups. It is possible to constrain the general location of the target fields based on sensitivity and systematic error considerations, however. Sensitivity models for the baseline design which include the effects of Zodiacal light and Galactic dust show that the preferred regions of sky are those close to the North ecliptic pole from February to July and close to the South ecliptic pole from August to January, and that the tightest constraints come from Galactic dust (see Sec.~\ref{sec:perf}. We can use relative sensitivity maps (Fig.~\ref{fig:sensmap}) to narrow down the optimum target field positions to between 15.5 and 16.5 hours Right Ascension and +55 and +60 degrees declination in the north and between 4 and 5 hours Right Ascension and -50 and -60 degrees declination in the south. Within each of these regions we expect to select 1 or 2 target fields, and each region will be observed for approximately 6 months before switching to the other region.

\subsection{Operational requirements}

In order to observe both northern and southern target fields under favourable Zodiacal light conditions we require an operational lifetime of at least 1 year, with a goal of 2 years or more. Based on this lifetime requirement and the results our initial power, thermal and communications budget analyses we have selected a nominal circular Sun-synchronous orbit at \SI{600}{\kilo\metre} altitude, \SI{96.7}{\minute} period and LTAN of 14:00.

\subsection{Observing constraints}
\label{sec:obsconst}

Useful scientific data can only be acquired when certain constraints are met. These include image sensor temperature (\SI{-40}{\celsius}) and minimum angular separations of the target field from the Earth's disc, the Sun and the Moon (\SI{25}{\degree}, \SI{60}{\degree} and \SI{60}{\degree} respectively). Preliminary analyses indicate that for the nominal orbit and approximate target field positions approximately 3--4 \SI{600}{\second} science exposures will be possible during each \SI{96.7}{\minute} orbit.

\subsection{Observing sequence}

The nominal sequence of operations for science observations is as follows:

\begin{enumerate}
\item{Slew spacecraft to target field, arcminute level pointing accuracy is sufficient. Roll angle is chosen to orientate the main solar array towards the Sun}
\item{Take a short exposure with the main image sensor to locate positions of pre-selected guide stars (2-3) on the image sensor, define regions-of-interest (RoIs) around them.}
\item{Begin continuous imaging of the guide star RoIs and start closed loop image stabilisation using guide star centroids (ADCS 2nd stage).}
\item{Take a long exposure using the remainder of the main image sensor. Usually this would be a 600s exposure but some shorter exposures will be taken to obtain unsaturated images of the brightest stars.}
\item{Repeat 4 until orbital motion causes observing constraints to be no longer satisfied, likely after 3--4 exposures}
\item{Perform Earth IR avoidance manoeuvres until observering constraints met again.}
\item{Repeat from 1 with pointing offset to the next dither position}  
\end{enumerate}

Over the course of 6 orbits Space Eye will obtain images at each of 6 dither positions in order to construct contiguous images for each of the 6 filters in the focal plane mosaic. Over the course of a 6 month period spent observing either the northern or southern targets the spacecraft will observe at roll angles spanning a range of \SI{180}{\degree} as the spacecraft tracks the Sun with its main solar array. This ensures that each position on the sky is observed at a range of field positions, which helps in suppressing residual calibration errors and internal reflections.

\subsection{Calibration sequences}

Space Eye will be extensively characterised before launch however these data will be supplemented with on orbit calibrations, including dark current measurements and flat fielding using a combination of Earth streak images, stellar photometry and median sky frames. The on-orbit measurements will account for changes caused by radiation damage, etc.  In addition science observations will be interspersed with exposures of bright stars (`PSF standards') to regularly characterise the faint, outer parts of the PSF wings in a strategy similar to that reported by Tujillo and Fliri\cite{Trujillo2015}.

\subsection{Data handling}

Each main image sensor image will be captured at 2000 x 1504 pixel resolution and 16 bits/pixel depth, i.e. each comprises \SI{6.016}{\mega\byte} of raw data. Image data will be stored as losslessly compressed Flexible Image Transport System (FITS) files, which typically have compression factors of around 2 for astronomical data. On board data processing will be limited to insertion of relevant spacecraft and instrument status information into FITS file headers, all raw images will be downlinked for processing and analysis on the ground. Based on the assumption of 3--4 science images per \SI{96.7}{\minute} orbit the average (compressed) data rate would be $\sim135$--\SI{180}{\mega\byte} per day. Downlinking this volume of data from a CubeSat using standard S-band communications and 1--2 ground stations would be a challenge, but is achievable\cite{Reisenfeld2015}.

\section{STATUS AND PLANS}

At the time of writing funding to complete the design, construction, testing, launch, and commissioning of the Australian Space Eye is being sought via the Australian Research Council's Linkage Infrastructure, Equipment \& Facilities (LIEF) national competitive grant scheme. 

The LIEF grant scheme allows consortia lead by an Australian higher education institution to seek part funding (typically $\sim50\%$) for a research infrastructure project from the government, with the remainder to come from the institutions of the consortium. The Space Eye LIEF consortium is lead by PI Lee Spitler of Macquarie University and includes astronomers, instrument scientists and engineers from 7 Australian universities (Macquarie University, Australian National University, UNSW, University of Sydney, University of Queensland, Western Sydney University and Swinburne), the Australian Astronomical Observatory, and both Tyvak and California Polytechnic State University, San Luis Obispo from the USA.

The outcome of our grant application is expected in October/November 2016 and, if successful, will be followed by a 3 year construction phase with launch and commissioning planned for H2 2019.


\acknowledgments 
 
This research made use of Astropy, a community-developed core Python package for astronomy\cite{Robitaille2013}, and the affiliated package ccdproc\cite{Craig2015}. Some of the results in this paper have been derived using the HEALPix\footnote{\url{http://healpix.sf.net/}}\cite{Gorski2004} package. This work also used the NumPy\cite{VanderWalt2011}, Scipy\cite{Jones2001}, Matplotlib\cite{Hunter2007} and IPython\cite{Perez2007} Python packages.

\bibliography{/home/ajh/Documents/Papers/library} 
\bibliographystyle{spiebib} 

\end{document}